\documentclass[aps,groupedaddress,superscriptaddress,prd, twocolumn]{revtex4}
\usepackage{hyperref}
\usepackage{amsmath,amssymb,amsbsy,amsfonts}
\usepackage{graphicx}
\usepackage{epstopdf}
\usepackage{bm}
\usepackage{soul}
\usepackage{array}
\usepackage{latexsym}
\usepackage{multirow}
\usepackage{sansmath}
\usepackage{subcaption}
\usepackage{lipsum}
\usepackage{float}
\usepackage{xcolor}
\usepackage{comment}
\usepackage{csquotes}	
\usepackage[plain]{algorithm}
\usepackage{algpseudocode}

\renewcommand{\theequation}{\arabic{section}.\arabic{equation}}

\def\be{\begin{equation}}
\def\ee{\end{equation}}
\def\bea {\begin{eqnarray}}
\def\eea {\end{eqnarray}}
\def\nn {\nonumber}

\makeatletter
\newcommand\makebig[2]{%
  \@xp\newcommand\@xp*\csname#1\endcsname{\bBigg@{#2}}%
  \@xp\newcommand\@xp*\csname#1l\endcsname{\@xp\mathopen\csname#1\endcsname}%
  \@xp\newcommand\@xp*\csname#1r\endcsname{\@xp\mathclose\csname#1\endcsname}%
}
\makeatother
\makebig{biggg} {3.0}
\makebig{Biggg} {3.5}
\makebig{bigggg}{4.0}
\makebig{Bigggg}{4.5}

\begin{document}

\title{Phase transitions due to Euclidean gravity}

\author{Mustafa Saeed} \email{mustafa.saeed@centre.edu}
\affiliation{Centre College,
Danville, KY, USA 40422}

\author{Diya Batool} \email{23100267@lums.edu.pk}
\affiliation{School of Science and Engineering, Lahore University of Management Sciences, Lahore 54792, Pakistan}

\author{Muhammad Muzammil} \email{muhammad.muzammil@unb.ca}
\affiliation{Department of Mathematics and Statistics, University of New Brunswick, Fredericton, NB, Canada E3B 5A3}

\author{Nomaan X} \email{nomaan.math@unb.ca}
\affiliation{Department of Mathematics and Statistics, University of New Brunswick, Fredericton, NB, Canada E3B 5A3}

\begin{abstract}
\vskip 0.2cm

We use Ising-like models to probe the thermal nature of Euclidean spacetime backgrounds. We determine which properties of the background -- curvature, the presence of a horizon, or temperature -- play a role in phase transitions. The geometries we use are Euclidean Schwarzschild, Rindler, extremal Reissner-N\"{o}rdstrom (ERN), Anti deSitter (AdS), and deSitter (dS). Among these, Rindler is flat, AdS does not have a horizon, and both AdS and ERN have zero temperatures. We find second-order phase transitions as the metric parameter is varied in all cases except for Rindler. Specifically, we find that the transition from order to disorder occurs as the curvature -- or Euclidean gravity -- increases. This supports our conjecture that Euclidean gravity is an essential ingredient for these phase transitions, as opposed to the presence of a horizon or temperature. Separately, since the selected geometries are position-dependent, the Ising-like models constructed on them are inhomogeneous, whereby they generalize the standard Ising model. We find that a consequence of this is that criticality does not correspond to maximal correlation lengths and scale invariance.

\end{abstract}
\maketitle
\setlength{\arrayrulewidth}{0.1mm}
\renewcommand{\arraystretch}{2}
\renewcommand{\theequation}{\arabic{equation}}

\hspace{\parindent}


\section{Introduction}
Boltzmann interpreted heat as the kinetic energy of microscopic particles, an interpretation later validated by quantum theory. Consequently, discoveries that a Schwarzschild black hole radiates like a blackbody with temperature proportional to the surface gravity of its horizon $T\sim \kappa$, and has an entropy proportional to its horizon area $S\sim A$, were interpreted as hints of an underlying quantum structure \cite{Haw71,Bek73,BarCarHaw73,Haw75}. This interpretation was supported by the finding that an accelerating observer in Minkowski spacetime -- who sees a horizon due to their acceleration $a$ -- registered a temperature $T\sim a$ \cite{Dav75,Unr76}. 

It has been proposed that these thermodynamic properties are due to the presence of horizons \cite{Pad05, Pad14, Pad19}, which can be physical or artificial depending on whether they can be eliminated by coordinate transformations. Thus, the black hole horizon is physical and the Rindler horizon is artificial, a coordinate transformation to an inertial reference frame yields a horizon-less, flat background.

Concerted efforts have been made to understand the link between thermodynamics and gravity \cite{Jac95, Wal99, Pad19}. These include statistical mechanical explanations for thermodynamic properties, both for black holes \cite{GibHaw77} and for Rindler backgrounds \cite{Laf87}. Such Euclidean quantum gravity approaches used the similarities between the statistical mechanics partition function at constant $T$ and the Euclidean path integral with imaginary time periodicity $T^{-1}$ \cite{Pol87}. 

They have led to the discovery of phase transitions in Euclidean gravity systems. Specifically, the Euclidean path integral applied to black holes in Anti deSitter (AdS) space -- the maximally symmetric solution to Einstein's equations with a negative cosmological constant $\Lambda$ -- leads to the Hawking-Page transition: for a fixed $\Lambda$, black holes of, and lesser than, a critical mass are unstable and decay into thermal AdS space while black holes more massive than the critical mass are stable against this decay \cite{HawPag83}. 

Moreover, general black holes with charge $Q$, called Reissner-N\"{o}rdstrom black holes, can be considered. For extremal Reissner-N\"{o}rdstrom (ERN) black holes, $Q=M$, the horizon is physical and $\kappa=0$. A na\"{i}ve comparison with the laws of thermodynamics suggests that $T=0$ and $S\sim A$. Indeed, it was found that their radiation spectrum is not thermal, whereby $T=0$ \cite{LibRotSon00, Goo20}. While the aforementioned Euclidean quantum gravity approaches are inapplicable here due to $T=0$, the $S\sim A$ relationship was demonstrated via string theory \cite{StrVaf96} and the Kerr-CFT correspondence \cite{GuiThoStr09}. It has also been shown that such black holes cannot be created classically without violating the weak energy condition \cite{BarCarHaw73}, which has fueled speculation that they are fundamentally quantum objects \cite{GibKal95}.

Of these thermal backgrounds, Schwarzschild has curvature, a physical horizon and $T\sim \kappa$; Rindler is flat and it has an artificial horizon with $T\sim a$; and the ERN black hole has curvature, a physical horizon with $T=0$. In addition to probing the thermal properties of these backgrounds further, we also test if a background with curvature, no horizon and $T=0$, like AdS, is thermal or not. Another motivation for selecting AdS is its utility in probing gravity-matter systems: due to its simplicity, AdS is useful in studying quantum fields \cite{Ish97} and in probing strongly coupled quantum field theories (QFTs) via the AdS-CFT conjecture \cite{Wit98}. 

For completeness, we also analyze the deSitter (dS) background, which is the maximally symmetric solution of Einstein's equations with a positive $\Lambda$. It has curvature and a horizon with $T\sim \sqrt{\Lambda}$ and entropy proportional to its area $S\sim A$ \cite{GibHaw77_2}. These thermodynamic features have statistical mechanics explanations due to Euclidean quantum gravity \cite{GibHaw77}. Separately, since a positive $\Lambda$ has a repulsive effect, the dS background is useful in the study of inflation. Consequently, quantum fields on dS backgrounds is an active area of research \cite{Allen:1985ux, Hollands:2010pr, SinGanPad13}. 

This work has two goals. Firstly, motivated by the utility of Euclidean metrics in gravitational path integrals \cite{GibHaw77}, we probe the thermal nature of these backgrounds. Our method of doing so is inspired by -- but distinct from the setup for -- the Hawking-Page transition which is caused by varying the black hole mass, a parameter in the metric. We investigate the thermal nature of a Euclidean background by determining how varying its metric parameter affects quantum matter. The formalism used is Euclidean QFT on curved backgrounds (QFTCB). It is different from standard QFTCB in which the background is Lorentzian \cite{BirDav82, Wal95, ParTom09}, and Euclidean QFT, which is done on a flat background with periodic time \cite{Sch58, Nak59, Sym69, Bel96, Sym04}. 

To keep the model simple, we discretize the background and restrict the field values to $\pm$ 1. These choices yield an Ising-like model on a curved Euclidean background in which the spin couplings are determined by the background. Thus, we probe the thermal nature of a Euclidean background by building an Ising-like model on it and determining whether varying the metric parameter causes the model to undergo a phase transition. We also identify which property of the background -- curvature, the presence of a physical horizon, or temperature -- plays a role in this phase transition (if any).

The standard Ising model is invaluable in studying interacting degrees of freedom in thermal equilibrium due to its simplicity and 
application to other systems which have similar symmetries and interactions \cite{Car96, Wil79}. Therefore the Ising model is used in diverse fields including condensed matter physics, biophysics \cite{Maj01, Zak22}, studies of complex networks like the internet \cite{Alb02}, quantum information \cite{Wu96, Bre17}, Euclidean QFTCB \cite{SaeHus24}, and quantum gravity \cite{Bon16, Bon24}. 

The Ising model comprises position-independent, or homogeneous, couplings. Since realistic systems are inhomogeneous, generalizations of the Ising model to include inhomogeneities are studied widely \cite{Yan761, Yan762, Wol81, Wol82, Cha18}. Such generalizations have lead to novel results regarding critical temperatures \cite{Yan762}, correlations \cite{Wol82} and even the absence of phase transitions at finite temperatures \cite{Wol81}. 

Since Euclidean backgrounds are position-dependent in general, the Ising-like models built on them will have position-dependent couplings. Therein lies our second goal: to introduce inhomogeneities in the standard Ising model in a controlled manner and explore the resultant generalizations. This may have applications in studies of spin models. 


This work extends \cite{SaeHus24} which found that Euclidean Schwarzschild and AdS black holes behave like heat baths for Ising spins. Specifically, it was discovered that decreasing the black hole mass causes the Ising spins to undergo a second-order phase transition from order to disorder. While the backgrounds used in \cite{SaeHus24} had curvature, physical horizons and $T\neq0$, we study and compare with geometries that are flat (Rindler), have no horizon (AdS), have an artificial horizon (Rindler) and have $T=0$ (ERN and AdS), to identify features of a Euclidean background that are essential for phase transitions. We also construct the Ising-like models differently to enable an improved interpretation and faster computations.

We find a second-order phase transition for the Schwarzschild case similar to the one found in \cite{SaeHus24}, indicating that the improved construction yields accurate results. Furthermore, we discover second-order phase transitions in all cases except for Rindler. For the Ising-like models on the curved Euclidean backgrounds, we find that the phase transition from order to disorder occurs due to an increase in curvature, or Euclidean gravity. Thus, a Euclidean background's temperature does not cause phase transitions, nor does the presence of a horizon; rather, the essential ingredient is Euclidean gravity. This differs from conventional wisdom regarding the role of temperature in causing phase transitions and the analogy between acceleration and the surface gravity of black hole horizons. 

We also discover that at criticality the model is not scale invariant, and that the correlation lengths are not the size of the lattice. This differs from similar results for the standard Ising model for which phase transitions are accompanied by scale invariance and maximal correlations. Thus, by introducing inhomogeneities we find results that are different from the standard case.


What follows is a review of the Euclidean backgrounds of interest in section II, construction of Ising-like models on them in section III, specifications of the simulation used and presentation of results in section IV, and reflections on the results in section V. 


\section{Euclidean backgrounds}
\label{EucBac}
This section reviews the relevant features -- curvature, the presence of a horizon, and natural temperature -- of the aforementioned backgrounds. To discuss curvature we will use the Kretschmann curvature scalar $K$ because it is -- unlike the Ricci scalar -- non-zero for vacuum solutions of Einstein's equations, like the black hole geometries we consider. By extension it is also useful in identifying curvature singularities, which we avoid. After reviewing the important features of a background, its Euclidean version will be derived. 

\subsection{Schwarzchild}
The metric of a Schwarzschild black hole is
\begin{align}
\label{eq:GenMet}
    ds^{2}=-F(r)dt^{2}+F(r)^{-1}dr^{2}+r^{2}d\Omega_{(2)}^{2}
\end{align}
where $F(r)=1-2Mr^{-1}$, $t\in\mathbb{R}$, $r\in\mathbb{R}^{+}$, and $d\Omega_{(2)}^{2}$ is the metric on a unit 2-sphere. The spacetime is curved as indicated by, for instance, the Kretschmann scalar: $K\sim M^{2}\,r^{-8}$. The metric and $K$ indicate a curvature singularity at $r=0$. Separately, the metric also indicates a coordinate singularity at $r=2M$. This is the horizon which we denote by $r_{0}$; it also provides a natural length scale for this geometry. Furthermore, it is a physical feature of the spacetime since it cannot be removed via coordinate transformations. At the horizon the surface gravity is $\kappa=\left(4M\right)^{-1}$ and the Kretschmann scalar is $K\sim M^{-6}$. 

Arriving at the Euclidean Schwarzschild metric requires a transformation to Euclidean time $\tau=it$ and a restriction to the Euclidean section (which in this case is the region with no curvature singularity); this gives
\begin{align}
ds^{2}=F(r)d\tau^{2}+F(r)^{-1}dr^{2}+r^{2}d\Omega_{(2)}^{2}
\end{align}
where $r\in[r_{0},\infty)$ and the curvature singularity is absent. The coordinate singularity can be removed by transforming to a new radial coordinate
\begin{align}
\label{eq:rsch}
    \rho=\kappa^{-1}\,\sqrt{F(r)}
\end{align}
in which the metric is
\begin{align}
\label{eq:EucSch}
ds^{2}=\left(\kappa\,\rho\right)^{2}d\tau^{2}+\left[\frac{r(\rho)}{2M}\right]^{4}d\rho^{2}+r(\rho)^{2}d\Omega_{(2)}^{2}, 
\end{align}
with $r(\rho)$ determined via \eqref{eq:rsch} and $\rho\in[0,4M)$. As $\rho\rightarrow 0$ the $\tau-\rho$ portion of the metric
\begin{align}
ds^{2}=\left(\kappa\,\rho\right)^{2}d\tau^{2}+d\rho^{2}
\end{align}
represents flat space in polar coordinates where $\kappa\,\tau$ and $\rho$ are the angular and radial coordinates respectively. The lack of periodicity 2$\pi$ in the former indicates a conical singularity at the origin. This is removed by imposing Euclidean time periodicity $2\pi \kappa^{-1}$. Thus, the Euclidean Schwarzschild metric is given by \eqref{eq:EucSch} with coordinate ranges $\tau\in[0,2\pi \kappa^{-1})$ and $\rho\in[0,4M)$. The inverse of the Euclidean time periodicity is the Hawking temperature $T=\left(2\pi \right)^{-1}\kappa = \left(8\pi M\right)^{-1}.$



\subsection{Rindler}
The second metric of interest is Rindler. It describes Minkowski spacetime from the perspective of a uniformly accelerating observer with acceleration $a$; thus for the Rindler geometry, $K=0$. The metric is 
\begin{align}
\label{eq:rindmet}
    ds^{2}=-\left(a\rho\right)^{2}dt^{2}+d\rho^{2}+dy^{2}+dz^{2},
\end{align}
where $t,y,z\in\mathbb{R}$, $\rho\in\mathbb{R}^{+}$, and $a$ is directed along $\rho$. A consequence of transforming to the observer's frame of reference is that $\rho \in \mathbb{R}^{-}$ is not accessible. Therefore, the Rindler metric has an artificial horizon at $\rho=0$ that is due to the observer's acceleration. A natural length scale for this geometry is $a^{-1}$.

In imaginary time, $\tau=it$, the metric is 
\begin{align}
   \label{eq:EucRind} ds^{2}=\left(a\rho\right)^{2}d\tau^{2}+d\rho^{2}+dy^{2}+dz^{2}.
\end{align}
The $\tau-\rho$ part represents flat space in polar coordinates where $a\tau$ and $\rho$ are the angular and radial coordinates respectively. Periodically identifying $a\tau$, whereby Euclidean time has periodicity $2\pi a^{-1}$, removes the conical singularity. Consequently, the singularity-free Euclidean Rindler background is given by \eqref{eq:EucRind} with coordinate ranges $\tau=[0,2\pi a^{-1})$ and $\rho\in\mathbb{R}^{+}$; the natural temperature is $T=\left(2\pi\right)^{-1}\,a.$ 

\subsection{Extremal Reissner-N\"{o}rdstrom}

Our third geometry is that of a charged black hole. 
This spacetime is described by the Reissner-N\"{o}rdstrom (RN) metric which has the form of \eqref{eq:GenMet} with $F(r)=1-2Mr^{-1}+Q^2r^{-2}$. 
The extremal case is when $M=Q$, for which the metric is
\begin{align}
    ds^{2}=-F_{e}(r)\,dt^{2}+F_{e}(r)^{-1}dr^{2}+r^{2}d\Omega_{(2)}^{2}
\end{align}
with $F_{e}(r)=\left(1-Mr^{-1}\right)^{2}\,\geq 0$. Extremal RN (ERN) has $K = 8M^{2}\left(7M^{2}-12Mr+6r^{2}\right)/r^{8}$. The metric and $K$ indicate a curvature singularity at $r=0$. Further, the metric indicates a coordinate singularity at $r=M$. This is the physical horizon $r_{0}$ for ERN, the natural length scale for this geometry. At the horizon, $\kappa=0$ and $K\sim M^{-4}$. 

In Euclidean time, $\tau=it$,
\begin{align}
    ds^{2}=F_{e}(r)\,d\tau^{2}+F_{e}(r)^{-1}dr^{2}+r^{2}d\Omega_{(2)}^{2};
\end{align}
since $F_e\geq0$, $r\in \mathbb{R}^{+}$ corresponds to the Euclidean section. Thus, Euclidean ERN (unlike Euclidean Schwarzschild) includes the portion of the spacetime inside of the horizon and the curvature singularity is still present. However, we only consider the Euclidean region outside the horizon for consistent comparisons with the Schwarzschild case. 

The coordinate singularity can be removed by transforming to a new radial coordinate (that does not have a $\kappa$ dependence like the Schwarzschild case, see \eqref{eq:rsch})
\begin{align}
\label{eq:EucERNrho}
    \rho=\sqrt{F_{e}}.
\end{align}
This gives the metric
\be 
\label{eq:EucERN}
\begin{split}
ds^{2}=&\,\rho^{2}d\tau^{2}+\bigg(\frac{1}{\rho^{2}}\bigg)\frac{M^{2}}{(1-\rho)^{4}}\,d\rho^{2}\\&\,+\bigg(\frac{M}{1-\rho}\bigg)^{2}d\Omega_{(2)}^{2}
\end{split}
\ee
with $\rho\in[0,1)$. To check if there is a conical singularity, it is convenient to transform to a new time coordinate $\bar{\tau}=\tau M$; then the $\bar{\tau}-\rho$ plane in the limit $\rho\rightarrow 0$ is
\be 
ds^{2}=\left(\frac{\rho}{M}\right)^{2}d\bar{\tau}^{2}+\left(\frac{M}{\rho}\right)^{2}d\rho^{2}.
\ee
This is 2d Euclidean AdS which does not have a conical singularity. Therefore the Euclidean ERN metric is given by \eqref{eq:EucERN} with $\tau\in \mathbb{R}$ and $\rho\in[0,1)$; the infinite periodicity in Euclidean time implies $T=0$. 

\subsection{Anti deSitter}
For AdS, our fourth background, we use the static form of the metric \cite{calvo2018geometry} which has the form \eqref{eq:GenMet} with $F(r)=1+r^{2}\,l^{-2}$. Here $l$ is the AdS scale, the natural length scale for this geometry, which is related to the negative cosmological constant by $l^{2}=-3\,\Lambda^{-1}$. The AdS geometry is negatively curved, as indicated by its Ricci scalar: $R \sim \Lambda$; its Kretschmann scalar is $K\sim \Lambda^{2}$.
Additionally, AdS does not have a horizon nor a singularity. 

To derive its Euclidean version, we transform to imaginary time $\tau=it$ to get
\begin{align}
ds^{2}=&\,\left[1+\left(\frac{r}{l}\right)^{2}\right]d\tau^{2}+\left[1+\left(\frac{r}{l}\right)^{2}\right]^{-1}dr^{2}\nn\\&\,+r^{2}d\Omega_{(2)}^{2}.
\end{align}
Since there is no conical singularity at the origin, Euclidean time periodicity is infinite which implies a vanishing temperature $T$. For consistent comparisons with the previous backgrounds we will use $\rho$ to label the radial coordinate of AdS instead of $r$. Therefore, the AdS metric is 
\begin{align}
\label{eq:Eucds}
ds^{2}=&\,\left[1+\left(\frac{\rho}{l}\right)^{2}\right]d\tau^{2}+\left[1+\left(\frac{\rho}{l}\right)^{2}\right]^{-1}d\rho^{2}\nn\\&\,+\rho^{2}d\Omega_{(2)}^{2}
\end{align}
with $\tau\in\mathbb{R}$ and $\rho\in\mathbb{R}^{+}$. 

\subsection{deSitter}
Our final geometry of interest is the dS spacetime. We consider the static patch of dS which allows for the definition of a future-directed time-like killing vector, and is relevant for field theory \cite{spradlin2002sitter}. The metric is of the form \eqref{eq:GenMet} with $F(r)=1-r^{2}\,l^{-2}$ where $l^2$ is related to the positive cosmological constant via $l^{2}=3\,\Lambda^{-1}$. The positive $\Lambda$ is associated with a repulsive gravitational effect whereby dS is used to model the accelerated expansion of the universe during inflation. The dS geometry is positively curved, as indicated by its Ricci scalar $R \sim \Lambda$. Further, its Kretschmann scalar is $K\sim \Lambda^{2}$. There is a coordinate singularity at $r=l$. This is also the dS horizon $r_{0}$. This horizon is an example of a cosmological event horizon; causal communication is impossible across it as the universe expands faster than the speed of light.


To Euclideanize this metric, we transform to Euclidean time $\tau=it$ and restrict to the region $r\leq l$ i.e within the static patch. For consistent comparisons with the black hole and Rindler backgrounds, where the horizons were at $\rho=0$, we transform the radial coordinate via $\rho=l-r$. This gives
\begin{align}
\label{eq:Eucds1}
ds^{2}=F(\rho)d\tau^{2}+F(\rho)^{-1}d\rho^{2}+(l-\rho)^{2}d\Omega_{(2)}^{2}
\end{align}
where $F(\rho)=\frac{\rho}{l}(2-\frac{\rho}{l})$ and the horizon is now at $\rho=0$. Near the horizon $\frac{\rho}{l}\rightarrow0$; in this limit the metric reduces to 
\begin{align}
ds^{2}=\left(\frac{2\rho}{l}\right)d\tau^{2}+\left(\frac{2\rho}{l}\right)^{-1}d\rho^{2}+l^{2}d\Omega_{(2)}^{2}.
\end{align}
The coordinate singularity can be removed by transforming to a new radial coordinate
\begin{align}
\label{}
    \tilde{\rho}=\sqrt{2l\rho}
\end{align}
in which the metric is
\begin{align}
\label{}
ds^{2}=\left(\frac{\tilde{\rho}}{l}\right)^{2}d\tau^{2}+d\tilde{\rho}^{2}+l^{2}d\Omega_{(2)}^{2}. 
\end{align}
This resembles flat space in polar coordinates in which the angular and radial coordinates are $l^{-1}\tau$ and $\tilde{\rho}$ respectively. To remove the conical singularity at the origin, we impose Euclidean time periodicity $2\pi\,l$. Thus, the Euclidean dS metric is given by \eqref{eq:Eucds1} with coordinate ranges $\tau\in[0,2\pi\,l)$ and $\rho\in[0,l)$. The inverse of the Euclidean time periodicity is the natural temperature $T=\left(2\pi l \right)^{-1}$

\subsection{Comparative analysis of the Euclidean geometries}
We now compare and review the important features of the backgrounds discussed. Firstly, all geometries except for Rindler are curved. Among these, the black hole backgrounds have position-dependent curvature: at the horizon, a natural length scale for these geometries, the curvature increases as $M$ decreases. This is indicated by the respective Kretschmann scalars $K$. Indeed, the same relationship holds at any distance that is a multiple of the horizon. Since the radial distance to the horizon is a natural length scale for the black hole backgrounds, the relationship between $K$ and $M$ captures how curvature depends on the geometry's metric parameter. For the AdS and dS spaces, the curvatures are constant; specifically $K\sim \Lambda^{2}$. Secondly, all backgrounds except for AdS have a horizon, with the Rindler horizon being artificial. Lastly, ERN and AdS have $T=0$, and the other backgrounds have a $T\neq 0$ which depends on their respective metric parameters. This discussion is summarized in table \ref{table:GeoFea}.

\begin{table}[H]
\begin{center}
\begin{tabular}{|p{1.2cm} |p{2 cm} p{1.2cm} p{1.2cm} p{1.2cm} p{1.2cm}|}
\hline
 & Schwarzschild & Rindler & ERN & AdS & dS\\
 \hline
$K$   & $M^{-6}$    & $0$ & $M^{-4}$ & $\Lambda^{2}$ & $\Lambda^{2}$\\
Horizon   & Present    & Present & Present & Absent & Present\\
$T$ &  $M^{-1}$  & $a$ & $0$& $0$ & $\sqrt{\Lambda}$\\
\hline
\end{tabular}
\end{center}
\caption{Features of the selected backgrounds. The top row shows the geometry's curvature, specified by the Krestschmann scalar, as a function of the metric parameter. For the black hole backgrounds, $K$ is evaluated at the horison $r_{0}$ or any multiple of it. For AdS and dS, $K$ is constant. The middle row specifies if a background has a horizon or not, and the last row indicates the dependence of a background's $T$ on its metric paramater. 
}
\label{table:GeoFea}
\end{table}

\section{Ising-like models on Euclidean backgrounds}
To define Ising-like models on these Euclidean backgrounds we introduce a massless scalar field on each of them. The Euclidean action for the scalar field on these geometries is discretized and its values are restricted to the set $\{-1,1\}$.

All the Euclidean metrics considered can be written as
\be 
\label{eq:genmet}
ds^{2}=u(\rho)d\tau^{2}+v(\rho)d\rho^{2}+w(\rho)ds_{(2)}^{2}
\ee
where $ds_{(2)}^{2}$ is the metric of the 2-manifold that is not of interest. The Euclidean action for a massless scalar field on a background metric $g_{ab}$ is 
\begin{equation}
\label{eq:sfa}
I_{E}\left[\Phi\right]= \frac{1}{2}\int d^{4}x\ \sqrt{g}\,  g^{ab}\partial_{a}\Phi\partial_{b}\Phi.
\end{equation} 
Since metric properties of interest are in the $\tau-\rho$ plane, the scalar field chosen obeys $\Phi\equiv\Phi(\tau,\rho)$.
With these assumptions the Euclidean action reduces to 
 \bea
\label{eq:conttodisc}
I_{E}\left[\Phi\right] 
&=& \frac{\sigma}{2} \int^\beta_0\int_0^{\bar{\rho}} d\tau\, d\rho \, w \nn\\
 &&\cdot\left(\sqrt{\frac{v}{u}}\, {\dot{\Phi}}^{2}+\sqrt{\frac{u}{v}}\,{\Phi'}^{2}\right)
\eea
where $\sigma$ is the area of the 2-manifold that is not of interest, and dots and primes denote partial derivatives with respect to $\tau$ and $\rho$. For the Schwarzschild, Rindler and dS cases, $\beta$ is the associated Euclidean time periodicity required to remove the conical singularity. For the ERN and AdS backgrounds, we choose a finite $\beta$ to ensure a finite lattice; a natural choice is Planck time whereby we select $\beta=1$. Thus, the Euclidean ERN and AdS backgrounds have $T=1$ \cite{Pol87}. Separately, the upper limit of the radial integration $\bar{\rho}$ must be lesser than the full extent of the radial coordinate. This avoids inclusion of an infinite amount of the background in the radial direction and, by extension, infinities in the Euclidean action. To select $\bar{\rho}$ we note that each background has a natural length scale associated with its parameter: for the black hole backgrounds it is their horizons $r_{0}$, for Rindler it is $a^{-1}$ and for AdS and dS it is their respective $l$. We select $\bar{\rho}$ in terms of these length scales, such that we include a significant portion of the background while remaining within an order of magnitude of the natural length scale. Considering this, we choose $\bar{\rho}=\rho(5r_{0})$ for the black hole backgrounds whereby for Schwarzschild $\bar{\rho}=8M/\sqrt{5}$ and for ERN $\bar{\rho}=4/5$. Analogously, we select 
$\bar{\rho}=5\,a^{-1}$ for Rindler, and $\bar{\rho}=5l$ for AdS. However, for dS we choose 
$\bar{\rho}=l$ for this is the full Euclidean section. Therefore, for all backgrounds except for ERN, $\bar{\rho}$ depends on the metric parameter. These choices, also summarized in table \ref{table:CooRan}, yield a discrete model on a finite lattice.
\begin{table}[H]
\begin{center}
\begin{tabular}{|p{0.4cm} |p{2 cm} p{1.2cm} p{1cm} p{0.7cm} p{0.7cm}|}
\hline
 & Schwarzschild & Rindler & ERN & AdS & dS\\
 \hline
$\beta$   & $8\pi M$    & $2\pi\,a^{-1}$ & 1 &1 & 2$\pi l$\\
$\bar{\rho}$ &  $8M/\sqrt{5}$  & $5\,a^{-1}$ & $4/5$& $5\,l$ & $l$\\
\hline
\end{tabular}
\end{center}
\caption{Coordinate ranges for the Euclidean backgrounds considered.
}
\label{table:CooRan}
\end{table}

For discretizing the action, consider the $\tau-\rho$ plane as a $N\times N$ lattice with spacing 
\be
\label{eq:dsin}
\epsilon_{\tau} = \frac{\beta}{N}, \quad \epsilon_{\rho} =\frac{\bar{\rho}}{N}.
\ee
Thus, the metric parameters which determine the $\tau$ and/or $\rho$ coordinate ranges, fix the discretization interval as well. However, regardless of the metric parameter the continuous background is replaced by a lattice with $N^{2}$ points. 

The chosen discretization is
\bea
&&\tau\rightarrow \tau_{m}=m\epsilon_{\tau},\qquad \rho \rightarrow \rho_{n} = n\epsilon_{\rho};\nn\\
&&f(\tau,\rho)\rightarrow f_{m,n},\nn\\
&&\dot{f} \rightarrow  \frac{f_{m+1,n}-f_{m,n}}{\epsilon_{\tau}},\nn\\
&&f' \rightarrow  \frac{f_{m,n+1}-f_{m,n}}{\epsilon_{\rho}}.
\eea
Lastly, the scalar field values are restricted to the set $\Phi=\{-1,1\}$. With this discretization, and the restricted values of the scalar field, the Euclidean action becomes the sum   
\begin{widetext}
\bea
\label{eq:imcb}
I_{E}\left[\Phi\right]&=&\sigma\sum_{m,n=1}^{N} 
\left\{-\Phi_{m,n}\left[w_{n}\,\sqrt{\frac{v_{n}}{u_{n}}}\,\left(\frac{\bar{\rho}}{\beta}\right)\,\Phi_{m+1,n}+w_{n}\,\sqrt{\frac{u_{n}}{v_{n}}}\,\left(\frac{\beta}{\bar{\rho}}\right)\,\Phi_{m,n+1}\right] \right. \nn\\
&&\left. +\,\Phi_{m,n}\,^{2}\left[w_{n}\,\sqrt{\frac{v_{n}}{u_{n}}}\,\left(\frac{\bar{\rho}}{\beta}\right)+\frac{1}{2}\left(w_{n}\,\sqrt{\frac{u_{n}}{v_{n}}}+w_{n-1}\,\sqrt{\frac{u_{n-1}}{v_{n-1}}}\,\right)\,\left(\frac{\beta}{\bar{\rho}}\right)\right] \right\}.
\eea
\end{widetext}
This is the desired Ising-like model, average values of which are computed using the Euclidean path integral
\begin{equation}
Z=\int{\mathcal{D}\Phi\, \exp\left(-I_{E}\left[\Phi\right]\right)}.
\end{equation}
In \eqref{eq:imcb} the first line has nearest neighbor interactions, and the second line has the self interactions. All interaction strengths are position-dependent. Further, 
the vertical and horizontal nearest neighbor interactions are different functions of the metric parameters. Thus, the metric parameter determines the interaction strengths, and whether the spins will be ordered or disordered. For this (spin--1/2) model, the self-interaction terms behave as external fields and do not influence the expectation values of the functionals of spins. 

The Ising-like model \eqref{eq:imcb} 
has a fixed number of spins. These are separated by discretizations $\epsilon_{\tau}$ and $\epsilon_{\rho}$ that depend on the metric parameter \eqref{eq:dsin}. This construction is different from the one in \cite{SaeHus24} which fixed discretization and varied lattice size according to the metric parameter. Therefore, in \cite{SaeHus24} different number of spins propagate on Euclidean backgrounds that differ by the value of the metric parameter. Further, this construction can lead to large lattices and unreasonable computational times. Thus, the new method we use has two primary advantages. Firstly, it enables the more appealing interpretation of the same number of spins probing a Euclidean background's thermality, regardless of the value of the metric parameter. Secondly, it facilitates fixing lattice sizes that enable fast computations. However, the cost is using lattices of different discretizations that differ by the value of the metric parameter. One of the backgrounds used in \cite{SaeHus24} was Euclidean Schwarzschild. Using the new method, we will compare our findings for the Schwarzchild case with those in \cite{SaeHus24} to ensure that both methods yield similar results. 



The Ising-like model \eqref{eq:imcb} was derived for metrics of type \eqref{eq:genmet}. By comparing the metrics for our selected geometries with \eqref{eq:genmet}, we derive the Ising-like models on those geometries. 

\subsection{Schwarzschild}
For the Euclidean Schwarzschild geometry, given by \eqref{eq:EucSch}, we have
\begin{equation}
\begin{split}
    u(\rho)=& \left(\frac{\rho}{4M}\right)^{2}, \quad
    v(\rho)= \left[\frac{r\left(\rho\right)}{2M}\right]^{4}, \\
    w(\rho)=& \,r\left(\rho\right)^{2}.
\end{split}
\end{equation}
For this geometry, $\sigma=4\pi$; therefore, the Ising-like model on a Euclidean Schwarzschild background is
\begin{widetext}
\begin{equation}
\label{eq:imsch}
\begin{split}
I_{E}\left[\Phi\right]=&\,16\pi M^{2}\sum_{m,n=1}^{N}\,\Biggg\{-\Phi_{m,n}\left[\frac{1}{\pi\sqrt{5}\,\left(\frac{\rho_{n}}{4M}\right)\left(1-\left(\frac{\rho_{n}}{4M}\right)^{2}\right)^{4}}\,\,\Phi_{m+1,n}+\pi\sqrt{5}\,\left(\frac{\rho_{n}}{4M}\right)\,\,\Phi_{m,n+1}\right]\\
&+\left[\frac{1}{\pi\sqrt{5}\,\left(\frac{\rho_{n}}{4M}\right)\left(1-\left(\frac{\rho_{n}}{4M}\right)^{2}\right)^{4}}+\frac{\pi\sqrt{5}}{2}\left(\left(\frac{\rho_{n}}{4M}\right)+\left(\frac{\rho_{n-1}}{4M}\right)\right)\right]\Phi_{m,n}\,^{2}\Biggg\}.
\end{split}
\end{equation}
\end{widetext}
Since the radial range depends on $M$ (see table \ref{table:CooRan}), factors of $\rho_{n}/4M$ that appear in \eqref{eq:imsch} take the same range of values regardless of $M$, whereby the phase transition is determined by the overall factor of $M^{2}$. Thus, decreasing $M$ is expected to weaken the interaction strengths, causing a phase transition from order to disorder. Further, since at the horizon -- or a fixed multiple of it -- $K\sim M^{-6}$ (see table \ref{table:GeoFea}), this phase transition is caused by an increase in Euclidean gravity.


\subsection{Rindler}
For Euclidean Rindler geometry, given by \eqref{eq:EucRind}, we get
\bea
u(\rho)=\left(a\rho\right)^{2},\quad v(\rho)=w(\rho)=1.
\eea
These and the choice $\sigma=1$ (i.e Planck area for the two-manifold that is not of interest) give the Ising-like model on  Euclidean Rindler space:
\begin{widetext}
\begin{equation}
\begin{split}
\label{eq:imri}
I_{E}\left[\Phi\right]=& \sum_{m,n=1}^{N} 
\Bigg\{-\Phi_{m,n}\left[\left(\frac{5}{2\pi} \right)\left(\frac{1}{a\, \rho_{n}}\right)\,\Phi_{m+1,n} + \left(\frac{2\pi}{5} \right) \left(a\,\rho_{n}\right)\Phi_{m,n+1}\right]\\ &\,+\,\Phi_{m,n}\,^{2}\left[\left(\frac{5}{2\pi} \right)\left(\frac{1}{a\, \rho_{n}}\right)+\left(\frac{\pi}{5}\right)\,\big(a\,\rho_{n}+a\,\rho_{n-1}\big)\right] \Bigg\}.
\end{split}
\end{equation}
\end{widetext}
Here, the radial range depends on $a^{-1}$ (see table \ref{table:CooRan}). Therefore the factors of $a\rho_{n}$ in \eqref{eq:imri} take the same range of values regardless of $a$. Since there is no overall factor of $a$, a phase transition is not expected as $a$ is varied.

\subsection{Extremal Reissner-N\"{o}rdstrom}
On Euclidean ERN, given by \eqref{eq:EucERN}, we have
\bea
u(\rho)&=&\,\rho^{2},\quad v(\rho)=\frac{M^{2}}{\rho^{2}\left(1-\rho\right)^{4}}, \nn\\ w(\rho)&=&\left(\frac{M}{1-\rho}\right)^{2}.
\eea
For this background $\sigma=4\pi$; thus the Ising-like model on Euclidean ERN is
\begin{widetext}
\bea
\label{eq:imern}
I_{E}\left[\Phi\right] &=& 4\pi M\sum_{m,n=1}^{N} 
\Bigg\{ -\Phi_{m,n} \left[\frac{4M^{2}}{5\,(1-\rho_n)^{4}\rho_{n}\,^{2}}\,\Phi_{m+1,n} + \frac{5\,\rho_{n}\,^{2}}{4}\,\Phi_{m,n+1} \right] \nn\\
&&+\,\Phi_{m,n}\,^{2} \left[\frac{4M^{2}}{5\,(1-\rho_n)^{4}\rho_{n}\,^{2}}+\frac{5}{8}\left(\rho_n\,^{2}+\rho_{n-1}\,^{2}\right) \right] \Bigg\}.
\eea
\end{widetext}
Since the radial range is independent of $M$ (see table \ref{table:CooRan}), factors of $\rho_{n}$ in \eqref{eq:imern} take the same range of values regardless of $M$. Thus, the phase transition is determined by the overall factor of $M$. Similar to the Euclidean Schwarzschild case, decreasing $M$ is expected to weaken the interaction strengths and consequently cause a phase transition from order to disorder. Additionally, since $K\sim M^{-4}$ at the horizon or any multiple of it (see table \ref{table:GeoFea}), this phase transition is caused by an increase in Euclidean gravity. 

\subsection{Anti deSitter}
For the AdS background, given by \eqref{eq:Eucds},
\begin{equation}
\begin{split}
u(\rho)=\left[1+\left(\frac{\rho}{l}\right)^{2}\right],\quad v(\rho)=u(\rho)^{-1}, \quad w(\rho)=&\rho^{2}.
\end{split}
\end{equation}
Here $\sigma=4\pi$; therefore the Ising-like model on Euclidean AdS is given by
\begin{widetext}
\bea
\label{eq:imads}
I_{E}\left[\Phi\right] &=& 4\pi\,l\sum_{m,n=1}^{N} 
\Bigg\{ -\Phi_{m,n} \left[5l^{2}\left(\frac{\rho_{n}}{l}\right)^{2}\,\left[1+\left(\frac{\rho_{n}}{l}\right)^{2}\right]^{-1}\,\Phi_{m+1,n} + \frac{1}{5}\,\left(\frac{\rho_{n}}{l}\right)^{2}\left[1+\left(\frac{\rho_{n}}{l}\right)^{2}\right]\,\Phi_{m,n+1} \right] \nn\\
&&+\,\Phi_{m,n}\,^{2} \left[5 l^{2}\left(\frac{\rho_{n}}{l}\right)^{2}\left[1+\left(\frac{\rho_{n}}{l}\right)^{2}\right]^{-1}+\frac{1}{10}\left(\frac{\rho_{n}}{l}\right)^{2}\left[1+\left(\frac{\rho_{n}}{l}\right)^{2}\right]+\frac{1}{10}\left(\frac{\rho_{n-1}}{l}\right)^{2}\left[1+\left(\frac{\rho_{n-1}}{l}\right)^{2}\right]  \right] \Bigg\}.
\eea
\end{widetext}
Since the radial coordinate range depends on $l$ (see table \ref{table:CooRan}), factors of  $\rho_{n}/l$ in \eqref{eq:imads} take the same range of values regardless of $l$, causing the phase transition to be determined by the overall factor of $l$. Therefore, decreasing $l$, or increasing $|\Lambda|$, is expected to weaken the spin interactions and cause a phase transition from order to disorder. Moreover, since $K\sim \Lambda^{2}$ (see table \ref{table:GeoFea}), this phase transition is caused by an increase in Euclidean gravity also. \\


\subsection{deSitter}
Lastly, for Euclidean dS given by \eqref{eq:Eucds1}
\bea
u(\rho)&=&\,\frac{\rho}{l}\left(2-\frac{\rho}{l}\right),\quad v(\rho)=u(\rho)^{-1}, \nn\\ w(\rho)&=&\left(l-\rho\right)^{2}=l^{2}\left(1-\frac{\rho}{l}\right)^{2}.
\eea
For this background, $\sigma=4\pi$. Thus, the Ising-like model on Euclidean dS is given by
\begin{widetext}
\bea
\label{eq:imds}
I_{E}\left[\Phi\right] &=& 4\pi\,l^{2}\sum_{m,n=1}^{N} 
\Bigg\{ -\Phi_{m,n} \left[\frac{\left(1-\frac{\rho_{n}}{l}\right)^{2}}{2\pi \left(\frac{\rho_{n}}{l}\right) \left(2-\frac{\rho_{n}}{l}\right)}\,\Phi_{m+1,n} + 2\pi \left(\frac{\rho_{n}}{l}\right)\left(2-\frac{\rho_{n}}{l}\right)\left(1-\frac{\rho_{n}}{l}\right)^{2}\,\Phi_{m,n+1} \right] \nn\\
&&+\,\Phi_{m,n}\,^{2} \Bigg[\frac{\left(1-\frac{\rho_{n}}{l}\right)^{2}}{2\pi \left(\frac{\rho_{n}}{l}\right) \left(2-\frac{\rho_{n}}{l}\right)}+ \pi\Bigg( \left(\frac{\rho_{n}}{l}\right)\left(2-\frac{\rho_{n}}{l}\right)\left(1-\frac{\rho_{n}}{l}\right)^{2}\nn\\
&&+ \left(\frac{\rho_{n-1}}{l}\right)\left(2-\frac{\rho_{n-1}}{l}\right)\left(1-\frac{\rho_{n-1}}{l}\right)^{2}\Bigg) \Bigg] \Bigg\}.
\eea
\end{widetext}
Here the radial range depends on $l$ whereby the range of values for $\rho_{n}/l$ in \eqref{eq:imds} is independent of $l$. Thus, the phase transition is determined by the overall factor of $l^{2}$. Consequently, decreasing $l$ -- or increasing $\Lambda$ -- is expected to weaken the interactions, causing the spins to undergo a phase transition from order to disorder. Additionally, since $K\sim \Lambda^{2}$ (see table \ref{table:GeoFea}), this phase transition is also caused by an increase in Euclidean gravity. 


\section{Numerical simulations and results}

While periodic Euclidean time is part of the Schwarzschild, Rindler and dS backgrounds, imposing periodicity in time for the ERN and AdS cases and in the radial directions for all backgrounds is computationally advantageous since it limits finite-size effects. However, due to periodicity in the radial direction, spins at the opposite ends of the radial coordinate range interact. This effect can be quantified through the ratio of spins at the radial boundaries to the total number of spins: $2N/N^{2}$. Since this effect decays as $2/N$, $N=50$ is chosen.



After selecting a particular Euclidean metric (Schwarzschild, Rindler, ERN, AdS or dS), its metric parameter, denoted $\zeta$, is fixed to some value. This determines $\epsilon_{\tau}$ and/or $\epsilon_{\rho}$. 
A perfectly aligned lattice is used as the starting configuration and subsequent configurations $x$ are generated according to the probability distribution $\displaystyle P(x)=\exp[-I_{E}(x)]$ using a Monte Carlo (MC) Markov Chain. For a MC step, a randomly selected spin is flipped, the resulting change in the action $\Delta I_{E}$ is calculated, and the spin flip is accepted if $\text{exp}\left(-\Delta I_{E}\right)\geq \text{rand}[0,1]$. A sweep is defined as $N^{2}$ steps and thermodynamic quantities such as alignment 
\begin{align}
A=\Bigg|\,\sum_{m,n=1}^{N}\Phi_{m,n}\,\Bigg|
\end{align}
and Euclidean action $I_{E}$ are calculated after a sweep. 


Lattice configurations are thermalized i.e they are generated till the plots of $I_{E}$ vs. number of sweeps represent normal thermal fluctuations about an average value. After thermalization, the average value of $A$ is computed with an additional $N_M=2000$ sweeps using the formula 
\be
\langle {A}\rangle = \frac{1}{N_M}\ \sum_{i=1}^{N_M} {A}_i.
\ee
Here $i$ is the sweep number and $A_{i}$ is the corresponding measurement of alignment. We also calculate the magnetic susceptibility which is given by
\begin{align}
    \chi=\frac{1}{T}\left(\langle {A^{2}}\rangle - \langle {A}\rangle^{2}\right).
\end{align}

\begin{figure*}[htbp!]
\centering
\includegraphics[width=1\linewidth]{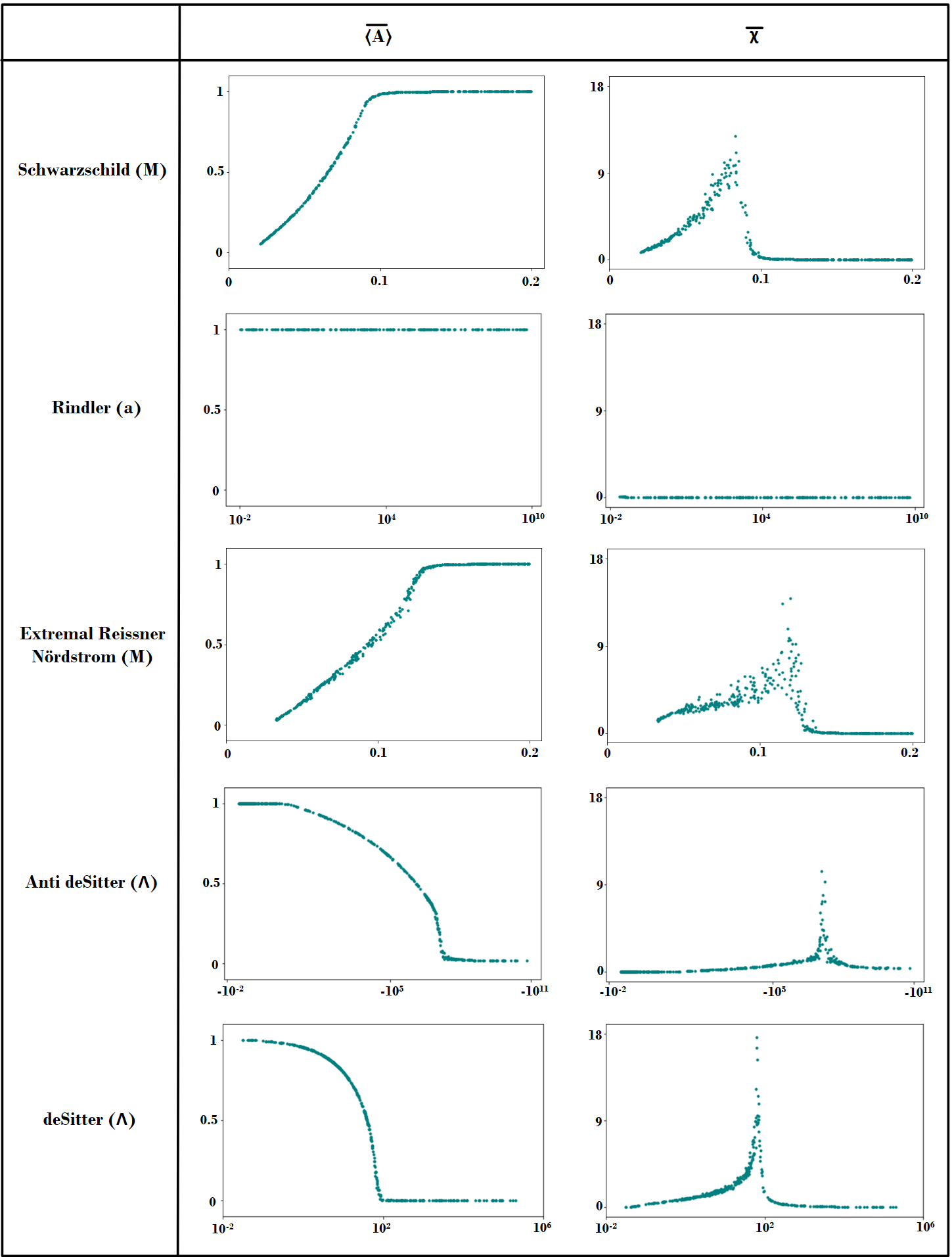}
\caption{\label{fig:f1}Plots of alignment per spin $\bar{\langle {A}\rangle}$ and magnetic susceptibility  per spin $\bar{\chi}$ vs. the metric parameter, specified in parenthesis next to the Euclidean background label.}
\end{figure*}

Here $T$ depends on the metric parameter for Schwarzschild, Rindler and dS cases but is a constant for the ERN and AdS backgrounds (see table \ref{table:GeoFea}). We further remove the effect of lattice size from these observables via
\begin{align}
    \bar{\langle A\rangle}=\frac{\langle A\rangle}{N^{2}}, \quad \bar{\chi}=\frac{\chi}{N^{2}}.
\end{align}
The process is then repeated for several values of $\zeta$. With sufficient data points, the plots for $\bar{\langle {A}\rangle}$ and $\bar{\chi}$ vs. $\zeta$ are generated for the selected background. The results are displayed in figure \ref{fig:f1}.

As argued previously, spins on the curved Euclidean backgrounds undergo a phase transition as the metric parameter is varied, but spins on Euclidean Rindler geometry do not undergo a phase transition (even for large $a$). Further, the transition from order to disorder occurs as the spins experience more Euclidean gravity. This supports our claim that Euclidean gravity is an essential ingredient for these phase transitions, as opposed to the presence of a horizon or natural temperature. 

For each of the curved background, the change in alignment is continuous and the magnetic susceptibility graph exhibits an approximate divergence at a critical value of the metric parameter $\zeta_{c}$. Thus, the phase transitions on the curved backgrounds are second-order and they occur at $\zeta_{c}$. Notably, in the geometrized units $G=c=\hbar=1$, the phase transitions occur at sub-Planckian $M$ for the black hole backgrounds and super-Planckian $\Lambda$ (corresponding to sub-Planckian $l$) for dS and AdS; the exact values are presented in table \ref{table:CriVal}. 
\begin{table}[H]
\begin{center}
\begin{tabular}{|p{0.4cm} |p{2 cm} p{1.2cm} p{1cm} p{1.2cm} p{0.8cm}|}
\hline
 & Schwarzschild & Rindler & ERN & AdS & dS\\
 \hline
$\zeta_{c}$   & 0.083    & NA & 0.12 & $-10^{7.1}$ & $10^{1.8}$\\
\hline
\end{tabular}
\end{center}
\caption{Critical value of the metric parameter $\zeta_{c}$
}
\label{table:CriVal}
\end{table}

Moreover, our result for the Schwarzschild background -- specifically, the second-order phase transition at sub-Planckian $M$ -- is similar to the one found in \cite{SaeHus24}. This  indicates that our new construction of Ising-like models on curved backgrounds not only gives accurate results, but it also has an appealing interpretation and requires inexpensive computations. These qualitative results are unchanged if a different $N$ is used. We choose $N=50$ because it produces accurate results in reasonable computational times. 

\begin{widetext}

\begin{figure*}[htbp!]
\centering
\includegraphics[width=\linewidth]{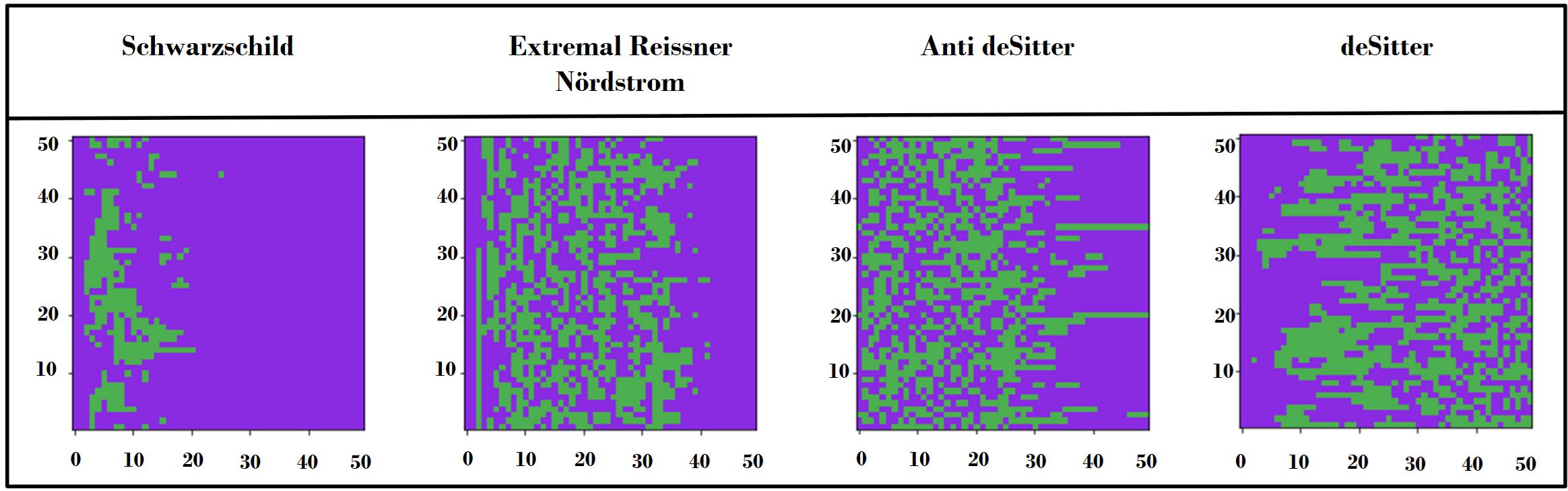}
\caption{\label{fig:f4}Thermalized lattices at criticality; purple (green) represents spins with values 1 (-1)}
\end{figure*} 

\end{widetext}

For the homogeneous Ising model, a phase transition corresponds to maximal correlation length and scale invariance. Figure \ref{fig:f4} shows thermalized Ising-like models on Euclidean backgrounds at criticality. For the black hole and dS cases, the left portions of the respective pictures represent spins at the horizons; for the ERN the horizon surface gravity is zero whereby the spins are ordered there. A common feature is that at criticality the correlation lengths are not maximal and the Ising-like model is not scale invariant. One natural follow up question is if this system will exhibit scale invariance for another value of $\zeta$, our answer for which is in the negative. Note that the Ising-like models we consider have inhomogeneous couplings. Further, at scale invariance -- as it is conventionally defined -- the spin behavior should be identical regardless of where on the lattice one zooms in. This conventional definition dictates that at a $\zeta$ at which the model becomes scale invariant, the couplings must become homogeneous; there is no value of $\zeta$ at which this occurs.

\section{Reflections}
We study the Ising model on various 4d 
Euclidean backgrounds. This incorporates ingredients of QFTCB \cite{BirDav82, Wal95, ParTom09} and Euclidean QFT \cite{Sch58, Nak59, Sym69, Sym04, Bel96} while being different from both: in QFTCB the backgrounds are Lorentzian, and in Euclidean QFT the absence of a conical singularity on flat space allows for time periodicity to be chosen due to other considerations. Our main purpose was to probe the thermality of Euclidean backgrounds by identifying what features are relevant for phase transitions in Ising spins.

We find that curvature, i.e Euclidean gravity, is an essential ingredient for phase transitions, as opposed to the presence of a horizon or natural temperature. 
Since curvature 
is an intrinsic property of the background and cannot be removed via coordinate transformations, these phase transitions are 
coordinate-independent results. 

Notably the results for the Schwarzschild and Rindler backgrounds differ. This, again, is due to curvature. Indeed, even if for the Schwarzschild background one focuses on the near horizon geometry and a small patch on the two-sphere -- whereby gravity is constant and linear -- the  geometry is still curved; specifically $K\sim M^{-6}$ (see table \ref{table:GeoFea}).
Thus, the phase transition is still expected for the case of constant linear gravity as $M$ is varied. 

Even though the spins depend on $\tau$ and $\rho$, the background geometry is 4d. Consequently, restricting the analysis exclusively to the $\tau-\rho$ plane is generally expected to influence the results. For example, in the black hole cases removing the 2-sphere affects the results due to the 2-sphere's radius being at least equivalent to the physical horizon, which depends on $M$. Thus, the 2-sphere is related to how $M$ appears in the Euclidean action, which is important for any phase transition. Similarly, removing the 2-sphere for the AdS and dS backgrounds results in a 2d space for which the relationship between $\Lambda$ and $l$ is ill-defined.  However, for the Rindler background, removing the $y-z$ part of the geometry does not change the results. This is because for the Rindler background, the $y-z$ components of the metric -- denoted $w$ -- are constant; specifically $w=1$ (see \eqref{eq:EucRind}). Given how $w$ appears in the Ising-like model \eqref{eq:imcb}, the $y-z$ part of the Rindler geometry has no affect on spin behavior, whereby removing it does not change the results.

Future avenues of exploration include probing other backgrounds via these Ising-like models. For instance, one can use near-extremal charged black holes for which $M\gtrsim Q$, and the geometry in extremal limit does not match that of ERN \cite{Car09}. In this context, one can probe near-extremal black holes using Ising spins and determine if the critical mass in the extremal limit matches the one found for ERN. Separately, one can study Ising spins on a Euclidean wormhole geometry and determine the size of the wormhole throat that maximizes the correlations between spins in the two asymptotically flat region. Work in this direction is in progress. Other future directions include replacing the spins by a scalar field that takes continuous real values over an interval, and adding self-interaction terms in the Euclidean action \eqref{eq:sfa}. 

Further, since the phase transitions on the curved backgrounds are second-order, one can compute critical exponents using techniques such as finite-size scaling \cite{Lan76,LanBin15,Pan04,Thi07}. These critical exponents can be used to determine the universality classes \cite{Wil79} for these models. However, such a calculation must be preceded by a careful selection of the independent parameter, since $T$ cannot be used in cases like ERN and AdS. \\

Lastly, one may view this work as a generalization of spin models on Euclidean space. This is because the coupling between spins here is position dependent and the metric parameter $\zeta$ affects vertical interactions differently from the horizontal ones. 
Already, the results of this work provide commentary on the known relationship between second order phase transitions, maximal correlation lengths and scale invariance. These three occur together at the critical temperature for the homogeneous Ising model. This work demonstrates that criticality does not coincide with maximal correlations and scale invariance for the Ising-like models on curved Euclidean backgrounds and that it is likely that there is no value of $\zeta$ at which the model becomes scale-invariant.


{\bf Acknowledgements} MS is supported by the FDC award at Centre College. MM is supported by NSERC. We thank Viqar Husain, Irfan Javed and Edward Wilson-Ewing for helpful comments on the manuscript. 

\bibliography{ref}

\end{document}